\documentclass[12pt]{article}
\pdfoutput=1
\usepackage{graphicx,color,hyperref,xspace,cite,mciteplus}

\newcommand{\registered}{\textsuperscript{\textregistered}\xspace}
\newcommand{\trademark}{\textsuperscript{\texttrademark}\xspace}

\begin{document}
\begin{flushright}
SLAC-PUB-15740\\
ANL-HEP-CP-13-47\\
FERMILAB-FN-0968-T\\
CERN-PH-TH/2013-217
\end{flushright}
\bigskip

\begin{center}
{\Large\bf Computing for Perturbative QCD\\[2mm] A Snowmass White Paper}
\end{center}

\medskip

\begin{center}
{\large Stefan H{\"o}che$^1$, Laura Reina$^2$, Markus Wobisch$^3$ (editors)\\
Christian~Bauer$^4$, Zvi~Bern$^5$, Radja~Boughezal$^6$, John~Campbell$^7$,
Neil~Christensen$^8$, Lance~Dixon$^1$, Thomas~Gehrmann$^9$, Junichi~Kanzaki$^{10}$, 
Alexander~Mitov$^{11}$, Pavel~Nadolsky$^{12}$, Fredrick~Olness$^{12}$, 
Michael~Peskin$^1$, Frank~Petriello$^{6,13}$, Stefano~Pozzorini$^9$, 
Frank~Siegert$^{14}$, Doreen~Wackeroth$^{15}$, 
Jonathan~Walsh$^4$, Ciaran~Williams$^7$} \\
\bigskip
$^1${\it %Theoretical Physics Department, 
SLAC National Accelerator Laboratory, Menlo Park, USA}\\
$^2${\it %Department of Physics, 
Florida State University, Tallahassee, USA}\\
$^3${\it %Department of Physics, 
Louisiana Tech University, Ruston, USA}\\
$^4${\it %Theoretical Physics Group, 
Lawrence Berkeley National Laboratory, Berkeley, USA}\\
$^5${\it %Department of Physics and Astronomy, 
University of California Los Angeles, Los Angeles, USA}\\
$^6${\it %High Energy Physics Division, 
Argonne National Laboratory, Argonne, USA}\\
$^7${\it %Theoretical Physics Department, 
Fermi National Accelerator Laboratory, Batavia, USA}\\
$^8${\it %Department of Physics and Astronomy, 
University of Pittsburgh, Pittsburgh, USA}\\
$^9${\it %Institute for Theoretical Physics, 
University of Zurich, Zurich, Switzerland}\\
$^{10}${\it KEK, %High Energy Accelerator Research Organization, 
Tsukuba, Japan}\\
$^{11}${\it %Theory Division, 
CERN, Geneva, Switzerland}\\
$^{12}${\it %Department of Physics, 
Southern Methodist University, Dallas, USA}\\
$^{13}${\it %Department of Physics and Astronomy, 
Northwestern University, Chicago, USA}\\
$^{14}${\it %Physikalisches Institut, 
Albert-Ludwigs-Universit{\"a}t, Freiburg, Germany}\\
$^{15}${\it %Department of Physics, 
State University of New York, Buffalo, USA}
\end{center}

\medskip

\section{Introduction}

One of the main challenges facing the particle-physics community to date
is the interpretation of LHC measurements on the basis of accurate and
robust theoretical predictions.  The discovery of a Higgs-like
particle in Summer 2012~\cite{Aad:2012tfa,Chatrchyan:2012ufa} serves
as a remarkable example of the level of detail and accuracy that must
be achieved in order to enable a
discovery~\cite{Dittmaier:2011ti,Dittmaier:2012vm,Heinemeyer:2013tqa}.
Signals for the Higgs boson of the Standard Model (SM) are orders of 
magnitude smaller than their backgrounds at the LHC, and they are 
determined by quantum effects.  Detailed calculations are therefore 
mandatory, and they will become even more necessary as we further 
explore the Terascale at the full LHC design energy.

Providing precise theoretical predictions has been a priority of the US 
theoretical particle-physics community for many years, and has seen an
unprecedented boost of activity during the last ten years. With the
aim of extracting evidence of new physics from the data, theorists
have focused on reducing the systematic uncertainty of their predictions
by including strong (QCD) and electroweak (EW) effects at higher orders
in the perturbative expansion. This is particularly important as
beyond Standard Model effects are expected roughly at the TeV scale. 
Typical decay chains of potential new particles would involve many 
decay products, several of which can be massive. The SM backgrounds 
are complex processes which call for highly sophisticated calculational 
tools in order to provide realistic predictions.

We have reached a time when no conceptual problems stay in
the way of breaking Next-to-Leading Order (NLO) perturbative QCD 
calculations into standard modular steps and automate them, 
making them available to the worldwide LHC community. 
The same is true for matching NLO calculations and parton showers.
It is implicit that automation will benefit greatly from a unified
environment in which calculations can be performed and data can be
exchanged freely between theorists and experimentalists,
as well as from the availability of adequate computational means
for extensive multiple analyses.

We nowadays see the frontier of perturbative
calculations for collider phenomenology being both in the
development and optimization of Next-to-Next-to-Leading Order (NNLO)
QCD calculations, sometimes combined with EW corrections, and in the
study of more exclusive signatures that requires resummation 
of logarithmically enhanced higher-order corrections to all orders.
It is also conceivable that techniques for matching NNLO fixed-order 
calculations to parton-shower simulations will be constructed in the 
near to mid-term future. In all cases, the availability 
of extensive computational resources could be instrumental
in boosting the exploration of new techniques as well as in
obtaining very accurate theoretical predictions at a pace and in a
format that is immediately useful to the experiments.

\section{Main results and recommendations}
This workshop provided a framework for implementing higher order  
calculations in a standardized computing environment made available 
by DOE at the National Energy Research Scientific Computing Center
(NERSC)~\cite{NERSC}.  Resource requirements were determined for the
calculation of important background and signal reactions at the
LHC, including higher order QCD and EW effects. Prototypical results 
have been summarized in a white paper~\cite{HPCWP}. Resource
requirements are also listed in Tab.~\ref{tab:summary}.

Different High Performance Computing (HPC) environments were tested
during this workshop and their suitability for perturbative QCD calculations 
was assessed. We find that it would be beneficial to make the national HPC 
facilities ALCF~\cite{ALCF}, OLCF~\cite{OLCF} and NERSC~\cite{NERSC} 
generally accessible to particle theorists and
experimentalists in order to enable the use of existing
calculational tools for experimental studies involving extensive
multiple runs without depending on the computer power and manpower
available to the code authors. Access to these facilities will also
allow prototyping the next generation of parallel computer programs
for QCD phenomenology and precision calculations.

The computation of NLO corrections in perturbative QCD has been automatized 
entirely. Resource requirements for NLO calculations determined during 
this workshop can thus be seen as a baseline that enables phenomenology 
during the LHC era. NNLO calculations are still performed 
on a case-by-case basis, and their computing needs can only be 
projected with a large uncertainty. It seems clear, however, that cutting edge 
calculations will require access to leadership class computing facilities.

The use of HPC in perturbative QCD applications is currently in
an exploratory phase. We expect that the demand for access to HPC
facilities will continue to grow as more researchers realize the 
potential of parallel computing in accelerating scientific progress. 
Long-term support of senior personnel, providing an interface 
between advanced computing research and application, may be 
required to fully exploit the potential of new technologies.
At the same time, we expect growing demand for educating young researchers 
in cutting edge computing technology. It would be highly beneficial 
to provide a series of topical schools and workshops related 
to HPC in HEP. They may be co-organized with experimentalists 
to foster the creation of a knowledge base.

Large-scale distributed computing in Grid environments 
may become relevant for perturbative QCD applications 
in the near future. This development will be accelerated if Computing Grids
can also provide access to HPC facilities and clusters where parallel 
computing is possible on a smaller scale. The Open Science Grid (OSG)~\cite{OSG} 
has taken first steps in this direction, and we have successfully used their
existing interface. The amount of training for new users could be minimized
if the OSG were to act as a front-end to the national HPC facilities
as well as conventional computing facilities.

\begin{table}
  \begin{tabular}{ccc}
    \hline
    Type of calculation & CPU hours per project & projects per year \\
    \hline\hline
    NLO parton level & 50,000 - 600,000 & 10-12\\
    NNLO parton level & 50,000 - 1,000,000 & 5-6\\
    Hadronic event generation & 50,000 - 250,000 & 5-8\\
    Matrix Element Method & $\sim$ 200,000 & 3-5\\
    Exclusive jet cross sections & $\sim$ 300,000 & 1-2\\
    Parton Distributions & $\sim$ 50,000 & 5-6\\
    \hline
  \end{tabular}
  \caption{Summary of computing requirements for the projects
    in Sec.~\ref{sec:requirements}
    \label{tab:summary}}
\end{table}

\section{Technology review}
Most state-of-the art perturbative QCD calculations are performed
using Monte-Carlo (MC) methods. The advantage of this technique
is the dimension-independent $1/\sqrt{N}$ convergence of the integral
(where $N$ denotes the number of sample points, or \textit{events}, 
used in the integration), and the possibility to generate events
with kinematics distributed according to the integrand
itself. A further advantage is that the calculation can be
parallelized trivially, by generating multiple sets of events with
different random seeds and combining them. This possibility is
exploited by experiments when running Monte Carlo event generators on
the Worldwide LHC Computing Grid~\cite{WLCG}. 

\subsection{Parallel computing using MPI}
Typical event generators need an initialization phase. During this
stage, the a priori weights of adaptive MC algorithms like
VEGAS~\cite{Lepage:1977sw} or the Multi Channel~\cite{Kleiss:1994qy}
are optimized for the particular calculation of interest. True
parallel computing on tightly coupled systems is highly advantageous 
in this phase in order to quickly exchange data needed for the 
optimization between different compute nodes.

The benefits of Message-Passing-Interface (MPI) parallelization
in this context has been assessed during this workshop by implementing
MPI communication into a representative Monte-Carlo event generation
framework. Similar strategies can be employed in other programs.  For
the purpose of this study we have chosen the parton-level event generator
BlackHat+Sherpa~\cite{Berger:2008sj,Gleisberg:2003xi,Gleisberg:2008ta}.
On the Cray XE6\trademark System ``Hopper'' at NERSC we observe 
strong scaling up to 1,024 cores. This makes it possible to attempt
calculations considered prohibitively time-consuming previously, 
and has been used in the computation of $pp\to W^\pm+$5~jets 
at NLO~\cite{Bern:2013gka}.

\subsection{Parallel computing using multi-threading}
In addition to MPI, multi-threading can be used to reduce the memory
footprint of executables on large-scale parallel computing systems, which
currently often suffer from a small memory per compute core. In this
case, executables can be designed such that various independent parts
of the calculation, like the evaluation of hard matrix elements and
the corresponding phase space, are performed in parallel.

Multi-threading can also be used efficiently in recursive algorithms
to compute matrix elements~\cite{Berends:1987me} and phase space 
weights~\cite{Byckling:1969sx}. This has been implemented and tested
in various event generators previously~\cite{Gleisberg:2008fv,Giele:2010ks}.  
In the context of this workshop we found that parallelization using MPI 
can be more efficient than multi-threading, as it scales with the number 
of nodes, while multi-threaded applications are limited by the number of 
compute cores on a single machine.  Hybrid approaches are very promising.

\subsection{Parallel computing using accelerators}
Several algorithms have been proposed, which allow to accelerate the
calculation even further using
GPUs~\cite{Hagiwara:2009aq,Giele:2010ks,Hagiwara:2013oka}.
Proof-of-concept implementations of these methods have shown great
potential and may be used on a larger scale in the near future.

\begin{table}
  \centering
  \begin{tabular}{ccc}
    \hline
    n-gluons  & integration &  generation\\
    & (BASES) & (SPRING) \\\hline\hline
    0 &             95  &                24\\
    1 &             84  &                44\\
    2 &             67  &                70\\
    3 &             39  &              $>$1000\\
    4 &             18  &                n.a.\\
    \hline
  \end{tabular}
  \caption{Ratios of total execution times of CPU and GPU programs 
    for MC integration and parton-level event generation in 
    $u\bar{d}\to W^++n$gluons with BASES/SPRING~\cite{Kawabata:1995th}.
    We used an NVidia Tesla C2075 GPU with CUDA\trademark 4.2 and
    an Intel\registered Core i7 2.67 GHz. 
    \label{tab:gpu}}
\end{table}

For the purpose of this study we use a GPU-implementation of a
$u\bar{d}\to W^++n$ gluons tree-level calculation~\cite{Hagiwara:2013oka}
as the benchmark process. Table~\ref{tab:gpu} shows ratios of
total execution times between CPU and GPU programs for MC integration 
and parton-level event generation with BASES/SPRING~\cite{Kawabata:1995th}.
We parallelized the event generation in the sense that multiple 
phase-space points are produced at the same time, in a load balanced
approach between the different CUDA kernels. The CPU program was not 
parallelized, hence the numbers shown in Tab.~\ref{tab:gpu} 
are indicative only of the gain when using GPUs alone.
Newer GPU architectures should further enhance performance.  

We have also tested MPI communication on an Intel\registered 
Xeon Phi\trademark coprocessor, which was made available to us by CERN 
OpenLab~\cite{COLA}. The clock frequency of its compute cores was 1.238~GHz, 
and the total memory of the system was 16~GB. The total number of compute 
cores was 244 (61$\times$4). We find that the execution speed of the 
executable can be reduced, depending on the complexity of the problem.
In simple MPI mode we observe strong scaling up to about 32 coprocessor 
cores. The ratio of computation time when run on a single coprocessor 
core, compared to a single CPU core ranges from 4.67 for $e^+e^-\to 2$jets 
over 7.52 for $e^+e^-\to 4$jets to 19.2 for $pp\to W^\pm+5$jets. 
It therefore seems more promising to use the coprocessor in offload 
and multi-threaded mode, rather than as an additional many-core processor. 
Current limitations in this context are entirely due to the structure 
of our executable, which has not yet been optimized for coprocessors. 
Substantially higher gain may therefore be expected in the future.

Developments for the application of GPUs and coprocessors to more complex 
and time consuming problems are ongoing.  Massively parallel computations 
using accelerators will likely become an important technique for perturbative 
QCD calculations and should be combined with other HPC methods.

\subsection{Suitability of existing HPC resources}
During this workshop, several HPC resources under supervision of the
U.S.\ Department of Energy (DOE) were made available for tests of
automated NLO calculations and Monte-Carlo event generators.  We have
benchmarked code performance at the following three facilities:
\begin{itemize}
\item[-] Cray XE6\trademark ``Hopper'' at NERSC~\cite{NERSC}\\
  24 AMD Opteron\trademark 2.1 GHz cores per node (153,216 total cores)\\
  32/64 GB RAM per node (6,000/384 nodes)\\
  Cray Gemini 3D Torus Network
\item[-] Cray XK7\trademark ``Titan'' at OLCF~\cite{OLCF}\\
  16 AMD Opteron\trademark 2.2 GHz cores per node (299,008 total cores)\\
  32 GB RAM per node (all nodes)\\
  NVidia\registered K20 GPU accelerators (18,688 total GPUs)\\
  Cray Gemini 3D Torus Network
\item[-] IBM\registered BlueGene\registered/Q test system ``Vesta'' at ALCF~\cite{ALCF}\\
  16 1.6 GHz PowerPC\registered A2 cores per node (32,768 total cores)\\
  16 GB RAM per node (all nodes)\\
  IBM 5D Torus Network
\end{itemize}

The two Cray systems resemble standard Linux environments,
which makes porting of existing codes convenient. Standard software
like the GNU compiler collection is available on all three systems. 
It was used for compiling our benchmark applications. It is to some 
extent simpler to test and debug code on the Cray architecture, 
as the environment includes interactive nodes which run a full fledged
Linux kernel.

We found that runtimes for the BlackHat+Sherpa event generator are identical 
to within 5\% on a Hopper node and on a Titan node. On these systems we have 
tested weak scaling up to 8,192 nodes and shown strong scaling up to 1,024 nodes.
These numbers are not necessarily representative for other applications, 
and they strongly depend on the process under consideration. 
It is likely that they will improve substantially over the next few years.

During our
tests on Vesta we encountered a larger I/O latency, which could be fatal for applications
designed to perform I/O operations on a per-node basis. The lower clock frequency 
of the IBM BlueGene\registered system does not allow direct comparisons between single 
compute cores on Vesta and the two Cray machines. In order to reduce turnover time, 
parallel codes would be favored strongly. To achieve similar performance 
in our benchmark, the number of compute cores had to be increased by 
a factor 2.2 compared to Hopper and Titan. However, the large number of cores 
on the test system Vesta and the corresponding production system Mira (786,432 total cores) 
may compensate this. We have tested weak scaling up to 16,000 cores on Vesta. 

\subsection{Suitability of the Open Science Grid}
We have tested the performance of Monte-Carlo event generation frameworks
(both parton-level and particle level) on the Open Science Grid~\cite{Altunay:2010zz,OSG}.
We found excellent usability, combined with very strong user support. 
The Condor-based glidein Workflow Management System, used by OSG, 
proved to be a very convenient tool for Monte Carlo event production
as larger jobs can easily be split into multiple subjobs.

In this manner we have carried out a Standard Model background 
study for Supersymmetry searches in the phenomenological MSSM within several days. 
The project required the use of 150,000 CPU hours and produced $\sim$1~TB of data. 
We used custom scripts to transfer the data from each worker node to the 
Storage Element (SE) at the site where the job ran. Later we transferred 
the data from the SEs to SLAC. We have started looking into using the 
OSG public storage service to automate this data handling.

We have explored MPI parallelization on the OSG by running small-scale 
HPC jobs. The number of cores accessible through the system is currently 
limited to the maximum number of cores per node (between 4 and 64). 
It will be highly beneficial to break this limitation in the future and
provide access to larger HPC facilities through the OSG as a common
interface.  This will reduce the training requirements for new
researchers entering the field, who want to make use of both,
large-scale distributed computing as well as HPC.

\section{Resource requirements}
\label{sec:requirements}

One of the aims of this workshop has been to provide a
quantitative estimate of the computational resources needed to
continue and expand the scientific program of the theory community
working at providing phenomenological predictions for the LHC
experiments. In this section we summarize the results obtained in
this context. The discussion focuses on three main building
blocks: higher order QCD and EW corrections to the parton level
cross sections, parton distribution functions, and event generators.

\subsection{Higher-order perturbative calculations}
\label{sec:nlo}

Collider events with large final-state particle multiplicity, and in particular 
with many jets, constitute the main backgrounds to many new physics searches.  
The need to describe such events with the best possible theoretical precision
has triggered substantial improvements in NLO calculations. 
On the one hand, Monte-Carlo programs have become available, which can compute 
all but the virtual corrections at NLO automatically and generate parton-level events. 
On the other hand, ``One-Loop Providers'' (OLP) have emerged, programs that 
efficiently compute the one-loop amplitudes entering the virtual corrections~\cite{
Berger:2008sj,Hahn:2000jm,Cullen:2011ac,Hirschi:2011pa,Cascioli:2011va,Reina:2011mb}. 
Combination of the two developments yields powerful and fully automated tools 
for LHC phenomenology. Hence, a standard was proposed in 2009 to combine MC and OLP 
in a unified manner~\cite{Binoth:2010xt}, which is used by all major projects.

\begin{table}
  \centering
  \begin{tabular}{llrr}
    \hline
    Process & Ref. & \multicolumn{2}{c}{Requirements}\\
    & & CPU [core~h] & Storage [GB] \\
    \hline\hline
    $pp\to W^\pm+5 jets$ & \cite{Bern:2013gka} & 600,000 & 1,500 \\
    $pp\to W^\pm+4 jets$ & \cite{Berger:2010zx} & 100,000 & 200\\
    $pp\to Z+4 jets$ & \cite{Ita:2011wn} & 200,000 & 200\\
    $pp\to Z+3 jets$ & \cite{Berger:2010vm} & 50,000 & 100 \\
    $pp\to 4 jets$ & \cite{Bern:2011ep} & 200,000 & 150\\
    \hline
  \end{tabular}
  \caption{CPU and Storage requirements for calculations on the
    list of important processes identified during the LesHouches
    series of workshops~\cite{AlcarazMaestre:2012vp}.
    Numbers assume cross-checks using at least
    two independent runs to guarantee the correctness of results
    and are reported for AMD Opteron\trademark processors running at 2.1~GHz.
    Storage requirements are reported for Root NTuple files which can be used
    to replicate the entire event analysis.
  \label{tab:nlo_wishlist}}
\end{table}

Table~\ref{tab:nlo_wishlist} shows CPU and storage requirements for 
typical NLO calculations required for LHC phenomenology. The numbers assume
at least one cross-check of the result to be performed with an entirely
independent run, and they include all parts of the NLO calculation.
The calculation is required to reach a Monte-Carlo uncertainty which makes 
the comparison with experimental measurements meaningful, see the
references for details. All numbers provided include cross section
calculations and the production of event samples, which are stored
in Root~\cite{Brun:1997pa} NTuple format and analyzed to produce the
histograms shown in the respective publications.

NLO accuracy has been recently reached also in techniques like the 
matrix-element method, which efficiently separates signal and background
events in searches at the LHC~\cite{Campbell:2012cz}.
The increased precision will greatly enhance the potential of analyses
focused on, for example, Higgs boson decays into $WW$ pairs, 
but it requires substantially larger computing power compared
to analogous LO techniques. Indeed, in addition to unobserved kinematic 
variables in the event, at NLO one must integrate over the extra degrees 
of freedom corresponding to emission of real radiation. 
Including integration over detector response functions and experimentally 
unobserved variables, this is equivalent to performing an entire 
cross section calculation for each event included in the analysis. 
Studying the existing $pp\to H\to WW$ candidate events in the 20 fb$^{-1}$ 
LHC data set requires 200,000 CPU hours in this approach.

\begin{table}
  \centering
  \begin{tabular}{llrr}
    \hline
    Process & Ref. & Requirements & CPU clock\\
    & & CPU [core~h] & [GHz]\\
    \hline\hline
    $pp\to W/Z$ & \cite{Melnikov:2006di,Li:2012wna} & 50,000 & 2.67 \\
    $pp\to H$ & \cite{Anastasiou:2005qj} & 50,000 & 2.67 \\
    $pp\to t\bar{t}$ & \cite{Baernreuther:2012ws,Czakon:2013goa} & 1,000,000 & 2.27\\
    $pp\to $ jets ($g$ only) & \cite{Ridder:2013mf} & 85,000 & 2.20 \\
    $pp\to H+$jet ($g$ only) & \cite{Boughezal:2013uia} & 500,000 & 2.67 \\
    \hline
  \end{tabular}
  \caption{Summary of computing requirements for NNLO calculations.
    Numbers were obtained on Intel\registered Xeon\registered CPU's with
    varying clock frequency and are therefore not directly comparable.
    \label{tab:nnlo_requirements}}
\end{table}

The precision being achieved in numerous benchmark measurements at the
LHC is imposing ever-increasing demands upon theoretical predictions.
NLO QCD calculations are no longer sufficient to match the level of
accuracy, for example, in measurements of W- and Z-boson properties.
Perturbative QCD calculations at the NNLO, sometimes combined with NLO
electroweak corrections, have become available in numerical programs
like FEWZ~\cite{Melnikov:2006kv,Gavin:2010az,Li:2012wna} and
FEHiP~\cite{Anastasiou:2005qj}, and will become available for several
other benchmark processes, including top-quark pair
production~\cite{Baernreuther:2012ws,Czakon:2013goa} and Higgs plus
jet production~\cite{Boughezal:2013uia} in the near future. A summary 
of current requirements is given in Tab.~\ref{tab:nnlo_requirements}.

Some programs like FEWZ are specifically designed to run in high
throughput mode on parallel computing systems. The NNLO QCD
corrections are thereby split into independent regions according to
their underlying singularity structure, and are integrated
independently on separate grids. In this manner, a full comparison
with measurements of the $d^2 \sigma/dM/dY$ distribution in
lepton-pair production divided into 150 bins requires approximately
50,000 CPU hours on Intel\registered Xeon\registered 2.67~GHz CPUs.
New subtraction schemes that provide a framework for NNLO calculations 
to arbitrarily-complicated processes also rely upon a splitting 
of the final-state phase space~\cite{Czakon:2010td,Boughezal:2011jf}.
 
Cutting-edge NNLO calculations that have recently become available, 
like top-quark pair production~\cite{Baernreuther:2012ws,Czakon:2013goa},
jet production~\cite{Ridder:2013mf} and Higgs plus
jet production~\cite{Boughezal:2013uia}, are much more demanding in
terms of computational resources, as can be seen in
Table~\ref{tab:nnlo_requirements}. Due to the increased demand for
NNLO predictions, we can expect substantially larger computing
resources to be needed in the near future. At the same time, the
development and adoption of standard techniques that
are optimized to provide public NNLO codes, is still very
preliminary. As a result, the numbers in
Table~\ref{tab:nnlo_requirements} have been obtained using different
methods and with the goal to reach a theoretical accuracy that
could efficiently compare with experiments. This makes the estimate
of needed computational resources, for the time being, very
process/calculation dependent. The access to HPC facilities will
shorten this exploratory phase and will allow a more rapid
convergence towards the selection of efficient techniques 
to be implemented in a broad range of future NNLO calculations.

While fixed-order perturbative QCD calculations can predict inclusive
jet observables very well, they cannot be applied to the analysis of 
exclusive jet bins due to the reduced accuracy associated with
a veto on the transverse momentum of real emissions.
Resummed calculations are required in this context, which have been
in the focus of interest recently due to their importance in the
$W^+W^-$ decay channel of the Higgs boson. They require computational
resources to run publicly available software for NLO and NNLO cross
section calculations in order to numerically extract coefficients needed
for the resummation, and also in order to match to the fixed-order result. 
Such studies have been performed for Higgs plus jet cross sections
on the Carver cluster at NERSC (Intel\registered Xeon\registered CPU at 2.67~GHz) 
and required approximately 300,000 CPU hours~\cite{Stewart:2013faa}. 
The demand for similar predictions is rising.

\subsection{Parton Distribution Functions}

Uncertainty in parton distribution functions (PDFs) constitutes the
leading theoretical uncertainty in many collider processes. It needs
to be reduced to realize the potential of multi-loop calculations for
QCD hard cross sections.

Computing needs of the future PDF analysis will be determined by a
number of trends.  First, implementation of fast NLO and NNLO
computations using the methods of ApplGrid~\cite{Carli:2010rw} 
and FastNLO~\cite{Kluge:2006xs} will speed up
PDF fits. These methods replace point-by-point $K$-factor lookup tables that
have been used in PDF fits to rapidly estimate higher-order radiative
contributions with some loss in the accuracy.  Without fast
computations or $K$ factor tables, the CPU time needed for fitting the PDFs
increases by 1-2 orders of magnitude.

Second, many numerical approximations in the computation of QCD cross
sections that speed up the current PDF analyses will need to be
eliminated in the future to match the accuracy of the NNLO and N$^3$LO
hard cross sections.

The global PDF analysis involves repetitive minimization, integration,
and interpolation in the space of many parameters.  The CPU time expenses
for every such step will need to be raised by up to an order of
magnitude, leading to nonlinear growth in the CPU time spent on the whole fit.

For example, in the current CTEQ and nCTEQ fits, scattering cross
sections are computed with numerical accuracy below a fraction of
percent. A typical recent study, such as production of
CT10 eigenvector sets, nCTEQ nuclear PDF sets, or investigation 
of charm mass dependence in the PDF analysis~\cite{Gao:2013wwa,
 Gao:2013xoa,Schienbein:2009kk}, required  3,000-10,000 CPU hours 
for 20-30 PDF parameters and using the $K$ factor
tables or fast (N)NLO cross sections. This time is spent on finding
the parametrizations of PDFs describing the data,
exploring the PDF parameter space in order to determine uncertainties
in the PDF parameters, and validating the results. Higher accuracy and
more free parameters and fitted processes can quickly increase the
CPU time demands, possibly by a factor 5-10 in the foreseeable future.

\subsection{Particle level event generators}

\begin{table}
  \centering
  \begin{tabular}{llllr}
    \hline
    Process & \multicolumn{2}{c}{$N_{jet}$} & Ref. & CPU [core~h] \\
    & NLO & LO & & \\
    \hline\hline
    $pp\to W^\pm+jets$ & $\le$2 & $\le$4 & \cite{Hoeche:2012yf} & 100,000 \\
    $pp\to h+jets$ & $\le$2 & $\le$3 & \cite{Hoeche:2013xxx} & 150,000 \\
    $pp\to t\bar{t}+jets$ & $\le$1 & $\le$2 & \cite{Hoeche:2013mua} & 250,000 \\
    $pp\to l\bar{\nu}\bar{l}'\nu'$ & $\le$1 & $\le$2 & \cite{Cascioli:2013gfa} & 50,000 \\
    \hline
  \end{tabular}
  \caption{Computing requirements for NLO-merged predictions in
    various benchmark processes, using the Sherpa event generator.
    Numbers assume cross-checks using at least two independent runs
    to guarantee the correctness of results.
    \label{tab:nlo_merging}}
\end{table}

Fully exclusive event generators~\cite{Buckley:2011ms} are a crucial tool 
to compare theoretical calculations to experimental observables including
realistic experimental cuts and detector resolution effects. 
There has been a large effort worldwide to increase the precision 
of the theoretical calculations these generators are based on, 
typically by including information from fixed order perturbation theory 
at NLO. The two matching methods MC@NLO~\cite{Frixione:2002ik} and 
POWHEG~\cite{Nason:2004rx}, which allow to combine NLO calculations with
parton showers, provide the theoretical basis. In addition, three different 
methods have been introduced recently to combine multiple NLO-matched 
calculations for varying jet multiplicity with each other.
They produce inclusive event samples, which can be reduced to NLO-accurate 
predictions at arbitrary jet multiplicity~\cite{Hoeche:2012yf,Lonnblad:2012ix,Frederix:2012ps}.
Such calculations are very demanding, since they rely on NLO calculations 
as an input to the simulation. The challenge of performing these NLO
calculations at high multiplicity can be appreciated by inspecting
Tab.~\ref{tab:nlo_wishlist}. Table~\ref{tab:nlo_merging} lists some 
exemplary computing requirements for the most challenging calculations
performed with MC event generators.

The Monte-Carlo framework Geneva~\cite{Alioli:2012fc} aims to go further 
and improve the formal accuracy of the resummation of large logarithms, 
which are typically only resummed at leading or next-to-leading logarithmic
order by the parton shower. Geneva also allows the combination of multiple 
NLO calculations with each other. The validation of the first results in 
$e^+ e^-\to$hadrons~\cite{Alioli:2012fc} required of the order of 200,000 
CPU hours. CPU time at the same order of magnitude is being used for the
validation of results in hadronic collisions~\cite{Alioli:2013vza}.

An important aspect in the construction of event generators is the validation
for new types of processes~\cite{Buckley:2010ar}, the preparation of 
public releases and their tuning~\cite{Buckley:2009bj}. 
All three aspects require substantial computing resources. Typically,
of the order of 50,000 CPU hours are spent for a full set of tests of the 
generator. The tuning process involves substantially more resources, typically
150,000 - 300,000 CPU hours.

\section{Conclusions}
\label{sec:summary}

Groups of particle theorists at both US universities and DOE
laboratories have been playing a leading role in each of the research
areas outlined in this report. To keep their impact and momentum at a time 
when the LHC is putting both precision SM studies and broad studies
of physics beyond the SM at high demand, these groups will need to have
access to extensive computing resources that could, to some extent, efficiently 
be provided within the framework of national supercomputing facilities.
At the same time, local resources must remain available for prototyping 
and testing of new applications.

We see two major benefits arising from the use of HPC facilities.  
First, a variety of existing
calculations/software can be made public within a common well-tested
framework, and in this way can be used for experimental studies
involving extensive multiple runs without depending on the computer
power and manpower available to their authors.  At the same time, new
sophisticated calculations can fully exploit the technological advantage
of the facility to provide cutting-edge results that would not otherwise 
be within reach. Specific examples of both uses have been given in this report.

Local computing resources provide a steady basis for small-scale
testing as well as prototyping and development of new calculations.
Efficient code development can only be guaranteed in an environment 
which allows unrestricted access to computing time, which is often
possible only with a resource partially or entirely under the management
of the researcher and its home institution. However, temporarily idle 
resources could efficiently be harvested by the Open Science Grid 
and thus made available to other researchers nationwide.

\section{Acknowledgments}

We are indebted to the Department of Energy, Office of Science, for 
initiating this study. We are especially grateful to Lali Chatterjee 
and Larry Price for stimulating discussions and for providing the 
required resources at NERSC. We thank Richard Gerber, Tom LeCompte 
and Richard Mount for their support in testing different LCFs.
We are indebted to Salman Habib for many stimulating and fruitful
discussions on code performance and optimization.
We thank Gabriele Garzoglio, Tanya Levshina and Marko Slyz for 
their help in using the OSG. We are grateful to Andrea Dotti 
and Pawe\l\ Szostek for help in testing the Intel Phi.

R. Boughezal is supported by the U.S.\ Department of Energy under Contract No. DE-\-AC02-\-06CH11357.  
J. Campbell and C. Williams are supported by the U.S.\ Department of Energy under Contract No. DE-\-AC02-\-06CH11359.
L. Dixon and S. H{\"o}che are supported by the U.S.\ Department of Energy under Contract No. DE-\-AC02-\-76SF00515.
A. Mitov is supported by ERC grant 291377 ``LHCtheory: Theoretical predictions and analyses of LHC physics: advancing the precision frontier''.
P. Nadolsky and F. Olness are supported by the U.S.\ Department of Energy under 
grant DE-\-FG02-\-04ER41299 and Early Career Research Award DE-\-SC0003870, and by the Lightner Sams Foundation.
F. Petriello is supported by the U.S.\ Department of Energy under Grant No. DE-\-SC0010143.
L. Reina is supported by the U.S.\ Department of Energy under grant DE-\-FG02-\-13ER41942.

\bibliographystyle{amsunsrt_modp}  
\bibliography{journal}

\end{document}